# Phase-controlled, heterodyne laser-induced transient grating measurements of thermal transport properties in opaque material

Jeremy A. Johnson, Alexei A. Maznev, and Keith A. Nelson

Department of Chemistry, Massachusetts Institute of Technology, Cambridge, Massachusetts

Mayank T. Bulsara and Eugene A. Fitzgerald

Department of Materials Science and Engineering, Massachusetts Institute of Technology, Cambridge, Massachusetts

T. C. Harman, S. Calawa, C. J. Vineis, G. Turner

Lincoln Laboratory, Massachusetts Institute of Technology

The methodology for a heterodyned laser-induced transient thermal grating technique for non-contact, non-destructive measurements of thermal transport in opaque material is presented. Phase-controlled heterodyne detection allows us to isolate pure phase or amplitude transient grating signal contributions by varying the relative phase between reference and probe beams. The phase grating signal includes components associated with both transient reflectivity and surface displacement whereas the amplitude grating contribution is governed by transient reflectivity alone. By analyzing the latter with the two-dimensional thermal diffusion model, we extract the in-plane thermal diffusivity of the sample. Measurements on a 5 μm thick single crystal PbTe film yielded excellent agreement with the model over a range of grating periods from 1.6 to 2.8 μm. The measured thermal diffusivity of $1.3 \times 10^{-6}$ m$^2$/s was found to be slightly lower than the bulk value.

# I. Introduction

Active research and development of innovative materials often creates the need for accurate characterization of their thermal transport properties[1]. Understanding how heat moves in thin films and nanostructured materials and finding ways to manipulate thermal transport hold promise in developing more efficient electronics as well as thermoelectric materials for solid state power generation. Indeed much progress in increased thermoelectric efficiency has been attributed to reduced thermal conductivity[2-5]. Introducing nanostructure such as layers, crystallite boundaries, or defects to scatter heat-carrying phonons and disrupt thermal transport is becoming an established method to reduce the lattice contribution to thermal conductivity.

A number of experimental approaches have been developed and utilized to measure thermal transport such as various ac calorimetric methods[6,7] including the 3ω technique[8] and optical methods including flash[9], pump-probe transient thermoreflectance (TTR)[10], and transient grating techniques[11-13]. Optical measurements of thin-film and other structured materials offer potential advantages over contact methods such as avoiding thermal contact[14] and interface conductance effects[15] and offering contactless, non-destructive characterization.

In the transient grating method (TG), also referred to as impulsive stimulated thermal scattering[16], two crossed laser pulses create a spatially sinusoidal temperature profile (thermal grating) in the sample and the dynamics of acoustic and thermal responses are monitored via diffraction of a probe laser beam. TG measurements have long proffered accurate and straightforward means for measuring acoustic and mechanical properties of complex samples, and commercial instruments exist for on-line measurements of layered materials[17,18]. In addition to ascertaining mechanical properties, TG measurements simultaneously allow determination of thermal transport properties[11,13,19-21].

In materials with weak optical absorption, thermal grating decay is exponential and the extraction of the thermal diffusivity from the data is straightforward. In opaque materials, the thermal grating is formed at the sample surface and a two-dimensional thermal diffusion equation needs to be solved to determine the temperature dynamics[21]. Furthermore, measurements on opaque samples are conducted in reflection geometry, with the diffracted probe beam intensity affected both by surface displacement and by variations of the reflection coefficient. The dynamics of the surface displacement and thermoreflectance are different[21] which further complicates the interpretation of the data. Oftentimes it is simply assumed that the diffraction signal is due to the surface displacement only[11], and we will show that this assumption can be highly inaccurate.

TG measurements can be greatly enhanced by optical heterodyne detection[22,23] which has become a standard technique adopted for many applications. Heterodyning allows separation of so-called phase and amplitude grating contributions to the transient grating signal[24,25]. In this work we utilize phase control to resolve ambiguities in the interpretation of transient thermal grating signals.

Below we introduce the relevant methodology for measuring thermal transport in an opaque sample using the transient thermal grating technique in reflection geometry with phase-controlled heterodyne detection. We isolate the thermoreflectance component of the signal, which allows direct comparison to a two-dimensional thermal transport model for the extraction of the thermal diffusivity. We utilize the method to measure the thermal diffusivity of a film of PbTe, an established material for thermoelectric applications[26].

**II. Amplitude vs. Phase Grating Signal**

In a reflection transient grating experiment as depicted in Fig. 1, two short excitation pulses with wavelength $\lambda_e$ are crossed at the surface of the sample of interest at an angle $\theta$. Optical interference leads to a periodic intensity profile with fringe spacing given by

$$L = \frac{2\pi}{q} = \frac{\lambda_e}{2\sin(\theta/2)} \ . \tag{1}$$

Absorption leads to spatially periodic heating and rapid thermal expansion, resulting in a transient grating with period $L$ and grating wavevector magnitude $q$. There are two main transient grating contributions from which a probe beam can be diffracted: a steady-state "thermal grating", of primary interest here, that remains until the heat diffuses away as well as an oscillating, standing-wave grating from impulsively generated surface acoustic waves (SAWs) that persists until the acoustic waves are damped out or simply leave the probing region[16]. A probe beam with wavelength $\lambda_p$ is incident on the transient grating. The diffracted light is spatially overlapped with an attenuated reference beam (derived from the same probe source as depicted in Fig. 1) and directed to a detector where the time-dependent intensity is measured.

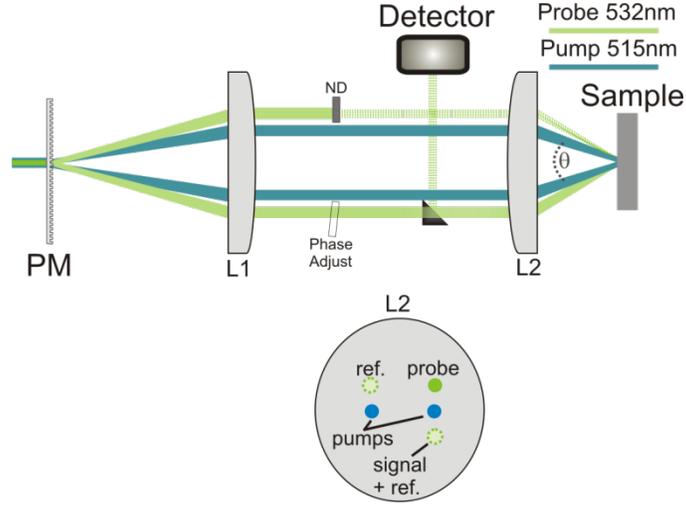

FIG. 1. (Color online) Schematic illustration of transient thermal grating experiment in reflection geometry. A diffractive optic, a binary phase-mask (PM), splits incident pump and probe beams into ±1 diffraction orders. The two resulting pump beams are focused and crossed at the sample surface by a set of lenses (L1 and L2), generating the transient thermal grating. Diffracted probe light is combined with an attenuated reference beam (ND) and directed to a fast detector. The relative phase difference between probe and reference beams is controlled by adjusting the angle of a glass slide (Phase Adjust) in the probe beam path. At the bottom, the spatial arrangement of beams on L2 is depicted.

Let us consider the details of the signal formation in the set-up described above. Optical fields associated with the probe and reference beams incident on the sample are approximated as plane waves respectively as

$$E_p = E_{0p} \exp\left(i\left(k_p^2 - q^2/4\right)^{1/2} z - i(q/2)x - i\omega_p t + i\phi_p\right) \quad (2)$$

and

$$E_R = t_r E_{0p} \exp\left(i\left(k_p^2 - q^2/4\right)^{1/2} z + i(q/2)x - i\omega_p t + i\phi_R\right) \quad (3)$$

where $E_{0p}$ is the initial probe amplitude, $k_p$ is the optical wavevector, $q$ is the transient grating wavevector magnitude defined above, $\omega_p$ is the optical frequency, $\phi_p$ and $\phi_R$ are the phases of probe and reference beams respectively, and $t_r$ is the attenuation factor for the reference beam. For a thin, time-dependent reflection grating and assuming the grating is a small perturbation, the complex transfer function, a combination of the complex reflection coefficient and phase change due to surface displacement, is given by

$$t^*(t) = r_0\left(1 + \cos(qx)\left[r'(t) + i\left(r''(t) - 2k_p u(t)\cos\beta_p\right)\right]\right) \quad (4)$$

where the dynamic complex reflection coefficient is given by $r^*(t) = r_0[1 + r'(t) + ir''(t)]$, $u(t)$ is the vertical surface displacement, and $\beta_p$ is the incidence angle of the probe beam on the surface. The first term in square brackets represents amplitude modulation of the incident field while the second term represents phase modulation. Assuming the sample is located at $z = 0$, the diffracted field can be obtained by multiplying the input field by the complex transfer function[27]. For the +1 diffraction order of the probe beam one obtains

$$E_{p(+1)} = \frac{1}{2} r_0 E_{0p} \left[ r'(t) + i\left(r''(t) - 2k_p u(t) \cos\beta_p\right)\right] \exp\left(-i\left(k_p^2 - q^2/4\right)^{1/2} z - i(q/2)x - i\omega_p t + i\phi_p\right) \quad (5)$$

and for the zero order reference beam

$$E_{R(0)} = r_0 t_r E_{0p} \exp\left(-i\left(k_p^2 - q^2/4\right)^{1/2} z - i(q/2)x - i\omega_p t + i\phi_R\right) a \quad (6)$$

The two beams are collinear and their interference gives an intensity

$$I_s = \frac{1}{2} I_{0p} R_o \left[ t_r^2 + r'^2(t) + \left(r''(t) + 2k_p^2 u^2(t) \cos\beta_p\right)^2 + 2t_r\left(r'(t)\cos\phi - \left(r''(t) - 2k_p u(t)\cos\beta_p\right)\sin\phi\right)\right] \quad (7)$$

where $I_{0p}$ is the intensity of the probe beam, $R_0 = |r_0|^2$ is the surface reflectivity, and $\phi = \phi_p - \phi_R$ is the heterodyne phase. In the absence of the reference beam, the non-heterodyned diffraction signal is given by

$$I_{non-het} = \frac{1}{2} I_{0p} R_0 \left[ r'^2(t) + r''^2(t) - 4r''(t) k_p u(t) \cos\beta_p + 4k_p^2 u^2(t) \cos^2\beta_p \right] \quad (8)$$

Thus without heterodyning the signal is comprised of a mixture of real and imaginary transient reflectivity and displacement terms. Since relative magnitudes of these terms are rarely known beforehand, unambiguous quantitative analysis of this signal would be difficult to say the least. If the reference beam intensity is much greater than that of the diffracted probe ($t_r >> r', r'', k_p u$), then the time-dependent signal is dominated by the heterodyne term

$$I_{het} = t_r I_{0p} R_0 \left[ r'(t) \cos\phi - \left(r''(t) - 2k_p u(t) \cos\beta_p\right) \sin\phi \right] . \quad (9)$$

If the time dependences of the reflectivity and displacement are different, the temporal shape of the signal will depend on $\phi$. By changing $\phi$, one can select the signal purely due to an amplitude grating at $\phi = 0, \pm\pi$ following the dynamics of $r'(t)$ and purely due to a phase grating at $\phi = \pm\pi/2$, representing the combination of $r''(t)$ and $u(t)$. Thus while it is straightforward to isolate the transient reflectivity signal $r'(t)$, it is not at all straightforward to isolate the displacement signal $u(t)$. Assuming that $r'(t)$ and $r''(t)$ follow the same temporal dependence, the displacement signal may be isolated at $\phi = -\arctan(r'/r'')$; however, this requires knowledge of the relative amplitudes of $r'$ and $r''$ which is not readily available[28]. Therefore unless the objective is determination of the surface modulation (for surface acoustic wave characterization, for example), the preferable strategy for the quantitative analysis of the TG signal is to set $\phi$ for pure amplitude grating and analyze the signal in terms of transient reflectivity rather than displacement.

**III. TG Experiment**

Heterodyne TG measurements were performed on a 5 μm-thick single crystal, undoped PbTe film grown via molecular beam epitaxy (MBE) on a (111) single crystal BaF$_2$ substrate. The deposition sequence entailed a 30-minute substrate bake at 575 °C to desorb any moisture or native oxide on the surface. After the substrate bake, the temperature was reduced to 325 °C and PbTe film growth was initiated with a 10:1 PbTe/Te flux. Thin film PbTe has shown promise for use in solid-state thermoelectric power generation and therefore its thermal transport properties, which may vary depending on fabrication conditions which affect defect density and morphology, are of interest.

In our current TG experiments, a short-pulsed laser beam ($\lambda_e$ = 515 nm, second-harmonic of HighQ femtoRegen, 60 ps uncompressed pulse width, 1 kHz repetition-rate) is split with a diffractive optic (binary phase mask pattern, PM) into two beams, which are then focused (300 μm 1/$e$ beam intensity radius, 0.66 μJ per pulse total energy) and crossed with angle $\theta$ at the surface of the sample of interest as depicted in Fig. 1. Pump-induced changes in the complex transfer function of the sample will lead to time-dependent diffraction of an incident continuous wave probe beam (532 nm, single longitudinal mode Coherent Verdi V5, 150 μm radius, 8.7 mW average power after chopping with an electro-optic modulator to 64 μs pulses to reduce sample heating). In the heterodyne detection scheme as described above, the diffracted signal is superposed with a reference beam that is derived from the same source and attenuated by a neutral density filter (ND), typically ND = 2.0. The superposed beams are directed to a fast detector (Hamamatsu C5658, 1 GHz bandwidth) and the time-dependent signal is recorded with an oscilloscope (Tektronix TDS 7404, 4 GHz bandwidth). The heterodyne phase $\phi$ is controlled by motorized adjustment of the angle of a glass plate in the path of the probe beam.

In a direct gap semiconductor like PbTe (0.31 eV bandgap at 300 K[29]), above-gap excitation results in fast thermalization of hot carriers, which promptly relax and transfer energy to the lattice[30,31]. In a TG experiment, energy is deposited with a sinusoidal spatial dependence at the sample surface, exponentially decaying with penetration depth 1/$\xi$. Heat will subsequently diffuse into the depth of the material as well as from grating peak to null parallel to the surface. Initial rapid thermal expansion will also launch counter-propagating surface acoustic waves resulting in a rapidly oscillating standing wave pattern. The combined effects of thermal and acoustic wave gratings will lead to a time-dependent complex transfer function described by Eq. 4.

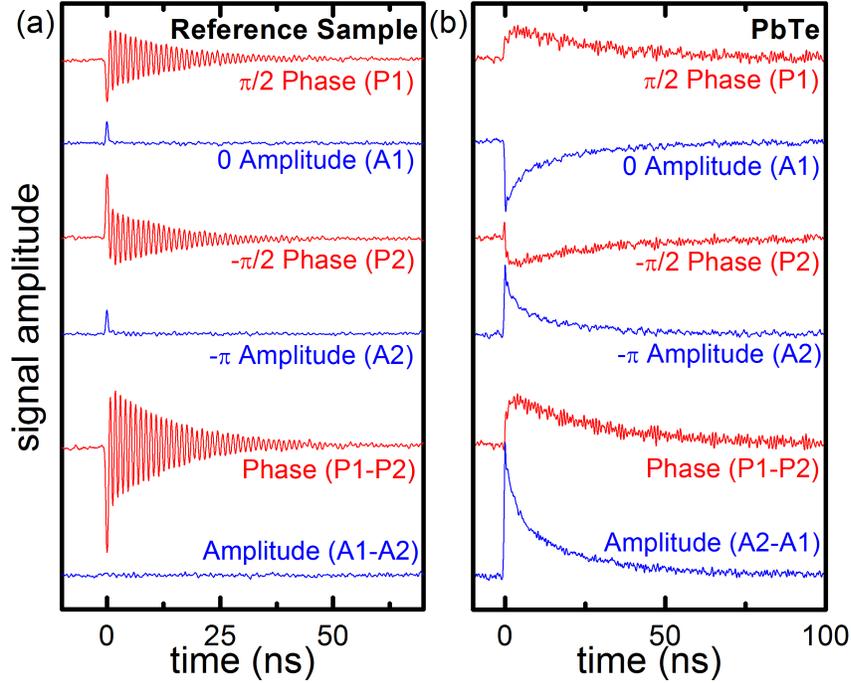

FIG. 2. (Color online) Transient grating data with a 2.05 μm grating period at four phase settings ($\phi = \pi/2$, 0, $-\pi/2$, $-\pi$) for (a) reference sample liquid m-xylene and (b) PbTe. In the transparent liquid reference sample, only the phase grating shows signal from damped acoustic waves. PbTe also shows acoustic waves only in the phase grating signal, whereas the amplitude grating dynamics are governed by the temperature decay only. Subtraction of signals corresponding to appropriate relative phase settings cancels spurious signals and increases signal-to-noise as shown in the lowest two traces for each sample.

To calibrate the heterodyne phase, we use a transparent liquid reference sample of m-xylene in transmission experimental geometry whose acoustic response yields a pure phase grating signal. It would be preferable to use an opaque reference sample that could be examined in the same reflection geometry as the PbTe sample; however, a number of materials were tried, including metals such as Au, Cu, Al and Ni, and semiconductors such as Si and GaAs, and all of them exhibited both phase and amplitude grating contributions at the probe wavelength of 532 nm. This underscores the point that the TG signal from opaque samples cannot generally be ascribed to the surface displacement alone.

A transparent sample in which the acoustic waves are generated via electrostriction (i.e., by impulsive stimulated Brillouin scattering[32]) ensures that the signal comes from the phase grating only. Switching between transmission and reflection geometries only requires a minor re-configuration of the detection set-up (the alignment of beams incident on the sample need not be changed), and separate detectors can be used for each scheme to allow for facile checking of the reference phase if it is desired.

m-Xylene transient grating signal traces were collected at four phase settings as depicted in Fig. 2(a). At $\phi = \pi/2$ and $\phi = -\pi/2$ we see decaying acoustic oscillations; the traces are nearly identical except of opposite sign, with the only difference due to a small amount of scattered pump light that reached the detector at $t=0$. The electrostrictively generated longitudinal acoustic waves modulate the density, which couples strongly to the index of refraction. This oscillating periodic index of refraction creates a time-dependent phase grating, and therefore when the acoustic oscillations are maximized, we know that the heterodyne phase is selected so signal is due to the phase grating signal contribution. When the acoustic oscillations are minimized, the signal is pure amplitude grating, and due to the absence of an amplitude grating signal in m-xylene, only a spike due to scattered pump light is observed at $t = 0$. Once the heterodyne phase is calibrated, we can take measurements from the PbTe sample and observe the pure amplitude or phase grating signal as shown in Fig. 2(b). In general, the complex reflectivity may include contributions from changes in temperature (thermoreflectance) as well as strain (photoelasticity). As is particularly apparent in the frequency content of the signals shown in the inset to Fig. 3, SAW oscillations are only present in the PbTe phase grating signal; the absence of the acoustic component in the amplitude grating signal indicates that $r'(t)$ reveals solely the thermoreflectance dynamics rather than the photoelastic response. As can be seen in Fig. 2, phase grating signals at $\phi = \pi/2$ and $\phi = -\pi/2$, as well as amplitude grating signals at $\phi = 0$ and $\phi = -\pi$, can be subtracted to cancel any spurious signals with no phase dependence. Examples of spurious signals include the scattered pump light at $t=0$ in the m-xylene traces and "non-heterodyned" signal components in Eq. 7. We also note that the difference between PbTe phase and amplitude grating signals is not limited to the acoustic oscillations that appear only in the former. The dynamics are totally different because the amplitude grating signal follows the thermoreflectance dynamics while the phase grating signal includes a combination of displacement and thermoreflectance dynamics. As will be shown in Sec. VI below, the initial rise of the phase grating signal is explained by the opposite signs of the displacement and thermoreflectance contributions.

Having established that maximum displacement signal indicates the pure phase grating in PbTe, traces with the heterodyne phase at $\phi=\pi/2,-\pi/2$, and $\phi=0,-\pi$ can be selected by respectively maximizing or minimizing the amplitude of SAW peaks in the signal Fourier transform. In practice this can be executed with good accuracy based on display of the signal Fourier transforms in real time on the oscilloscope. The accuracy can then be optimized through the following procedure to correct for the error in the heterodyne phase. Let us assume that the signals nominally corresponding to $\phi_1= 0$ and $\phi_2= \pi/2$ were in fact collected at $\phi_1= \delta_1$ and $\phi_2= \pi/2+\delta_2$, where $\delta_1$ and $\delta_2$ are unknown phase errors. According to Eq. (9), the respective signal waveforms are given by

$$S_1 = 2t_r I_{0p} R_0 \left[ r'(t)\cos\delta_1 - \left( r''(t) - 2k_p u(t)\cos\beta_p \right) \sin\delta_1 \right]$$
$$S_2 = -2t_r I_{0p} R_0 \left[ r'(t)\sin\delta_2 + \left( r''(t) - 2k_p u(t)\cos\beta_p \right) \cos\delta_2 \right] . \quad (10)$$

We find the corrected signal as a linear combination $S_1^* = S_1 + bS_2$, where the factor $b$ is adjusted to minimize the SAW Fourier peak in the spectrum of the corrected signal. The value of $b$ that makes the displacement contribution vanish is $\sin\delta_1/\cos\delta_2$, hence

$$S_1^* = 2t_r I_{0p} R_0 r'(t) \left[ \cos\delta_1 + \frac{\sin\delta_1 \sin\delta_2}{\cos\delta_2} \right] = 2t_r I_{0p} R_0 r'(t) \left[ 1 + \delta_1\delta_2 - \frac{\delta_1^2}{2} \right] \quad (11)$$

Thus the corrected signal $S_1^*$ recovers the pure amplitude grating response determined by $r'(t)$. The values of $b$ were in the range ±0.1, indicating that the final correction was small in all cases. For all four heterodyne phase settings, traces of 40,000 averages were collected for transient grating periods $L$ = 1.55, 1.80, 2.05, 2.40, and 2.75 μm. For the thermal transport analysis described in the following sections, we used the amplitude grating response obtained by subtracting corrected $\phi=0$ and $\phi=-\pi$ signals.

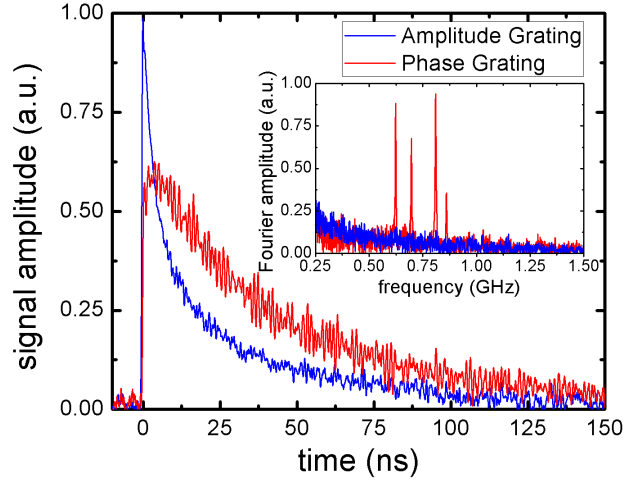

FIG. 3. (Color online) Amplitude and phase grating signals for a 2.05 μm grating period. The Fourier transform inset shows four clear thin film waveguide acoustic modes originating from the displacement contribution to the phase grating signal. The amplitude grating signal tracks temperature dynamics and is not influenced by the displacement.

## IV. Analysis of the Thermal Transport

In analyzing the thermal grating decay, we follow[21] but allow for a finite absorption depth of the excitation light. Temperature dynamics following transient grating heating at the surface of a semi-infinite sample are given by the thermal diffusion equation

$$\rho c \frac{\partial T}{\partial t} - k_x \frac{\partial^2 T}{\partial x^2} - k_z \frac{\partial^2 T}{\partial z^2} = Q_0 \cos(qx)\exp(-\xi z)\delta(t) , \tag{12}$$

where $x$ is the grating dimension, $z$ is the depth into the material, $T$ is the deviation from equilibrium temperature, $\rho$ is density, $c$ is heat capacity, $Q_0$ is the energy deposited by the short (δ-function) laser pulse to penetration depth $1/\xi$, and $k_x$ and $k_z$ are the thermal conductivity components respectively parallel and perpendicular to the sample surface, which are not assumed to be equal in order to incorporate the treatment of samples with anisotropic thermal transport. Appropriate boundary conditions are given by

$$\begin{aligned} \frac{\partial T}{\partial z}(z=0) &= 0 \\ T(z=\infty) &= 0 \end{aligned} . \tag{13}$$

Assuming a sinusoidal spatial dependence of $T$ on $x$ according to the source term, the solution to Eq. (12) is

$$T(z,x,t) = Af(z,t)\left[\cos(qx)\exp(-\alpha_x q^2 t)\right] , \tag{14}$$

where $A$ is the amplitude and $f(z,t)$ which accounts for thermal diffusion into the depth of the material is given by

$$f(z,t) = \frac{1}{2}e^{\zeta(-z+\alpha_z\zeta t)}\left[2 - \text{Erfc}\left(\frac{z - 2\alpha_z\zeta t}{2\sqrt{\alpha_z t}}\right) + e^{2\zeta z}\text{Erfc}\left(\frac{z + 2\alpha_z\zeta t}{2\sqrt{\alpha_z t}}\right)\right]. \quad (15)$$

Efrc($x$) is the complimentary error function defined as $\text{Efrc}(x) = \frac{2}{\sqrt{\pi}}\int_0^x e^{-t^2}dt$ and $\alpha_z$ is the cross-plane thermal diffusivity ($\alpha = k/\rho c$). Our interest is the temperature dynamics at the surface of the material, ($z=0$ where we probe the response), and so Eq. 14 can be simplified to

$$T(z=0,t) = Ae^{\alpha_z\zeta^2 t}\text{Efrc}\left(\zeta\sqrt{\alpha_z t}\right)\left[\cos(qx)\exp(-\alpha_x q^2 t)\right]. \quad (16)$$

In Eqs. 14 and 16, the simple expression in brackets, where $\alpha_x$ is the in-plane thermal diffusivity and $q$ is the transient grating wavevector given by Eq. 1, results from the solution to the 1-D heat equation with periodic source and accounts for thermal transport from transient grating peak to null parallel to the surface. This exponential decay, which is of primary interest to us, exhibits a decay rate that varies as $q^2$. To better understand the time domain decay, we consider the analytical solution of Eq. 14 in the limit of surface heating ($1/\zeta \rightarrow 0$). The temperature grating decay at the surface then follows the form

$$T(z=0,x,t) = \frac{A}{\zeta\sqrt{\pi}}(\alpha_z t)^{-1/2}\cos(qx)\exp(-\alpha_x q^2 t). \quad (17)$$

In this case, the transient grating measurement is insensitive to the cross-plane diffusivity $\alpha_z$ because changing it is equivalent to changing the amplitude $A$ of the signal. On the other hand, the in-plane thermal diffusivity contained in the exponentially decaying term can be determined by fitting the signal waveform to Eq. 17. In Fig. 4, we see illustrated the signal decay as modeled by Eq. 17 for three grating wavevectors. The heavy dark curve denotes a $t^{-1/2}$ decay and the next three curves show simulations with progressively smaller grating periods (larger grating wavevectors). As the grating period gets smaller, it takes less time for heat to diffuse from grating peak to null, and the signal decay is mainly due to the in-plane thermal diffusion described by $\alpha_x$. At larger grating periods as the decay gets longer, the signal is mainly due to the cross-plane thermal diffusion into the depth of the material and the $t^{-1/2}$ decay will start to dominate the signal, leading to less sensitivity to the in-plane diffusivity, as indicated by the dotted traces with ±20% values of the diffusivity.

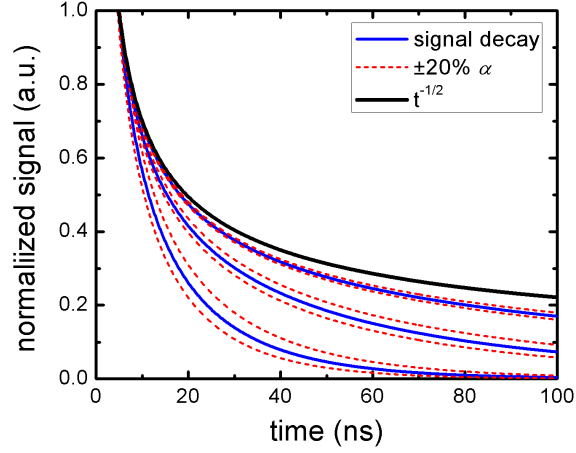

FIG. 4. (Color on-line) Simulation of signal illustrating sensitivity to the in-plane diffusivity $\alpha_x$ as a function of transient grating period. The heavy uppermost curve shows $t^{-1/2}$ decay that describes cross-plane thermal diffusion in the limit of surface heating. The next three solid curves are simulated traces for an in-plane diffusivity of $\alpha_x = 1.3 \times 10^{-6}$ m²/s for grating periods of 4.3, 2.1, and 1.1 µm. The dotted traces are simulations for ±20% values of $\alpha_x$. At short grating periods, the temperature grating decays mainly due to in-plane thermal diffusion, and the dashed curves show high sensitivity to the in-plane thermal diffusivity value $\alpha_x$. At longer grating periods the signal decays mainly through diffusion of heat into the bulk, and the signal becomes less sensitive to $\alpha_x$.

We note that Eq. 17 diverges at $t = 0$. This is not the case for the full solution given by Eq. 16 that accounts for the finite penetration depth of the laser heat source (~15 nm for 515 nm light in PbTe), which essentially smoothes out the diverging temperature over an initial time period $\Delta t = (4\alpha_z \zeta^2)^{-1}$. In principle, analyzing the initial part of the signal may help determine the cross-plane diffusivity. However, due to limits in temporal resolution and potential complications due to excited carrier diffusion, we fit the decay after 10 ns to extract only the in-plane diffusivity. At these delay times, the Eqs. 16 and 17 give indistinguishable results, so we fit out data to the simpler Eq. 17. Typical transient thermoreflectance (TTR) measurements with picosecond time resolution and a metal transducer layer[33] are more sensitive to the cross-plane diffusivity, so TTR and TG measurements could be used in tandem to reveal anisotropic thermal transport properties.

## V. Thermal Diffusivity Measurement

Fig. 5 shows amplitude grating signals for three representative grating periods along with the fits to Eq. 17, which show excellent agreement. The extracted thermal diffusivity was nearly identical for all data sets as shown in the inset to Fig. 5, giving an average value of 1.32(±0.03) × 10⁻⁶ m²/s. The fact that for a range of grating periods we get an excellent fit yielding the same value of $\alpha_x$ supports the soundness of our methodology. The determined thermal diffusivity is lower than the published value for bulk single-crystal PbTe of 1.8 × 10$^{-6}$ m$^2$/s at room temperature[33], but we note that a PbTe thin film deposited on BaF$_2$, although single crystalline, is prone to a high dislocation density (>10$^8$ cm$^{-2}$ determined for similar samples from cross-sectional transmission electron microscopy images) due to the lattice parameter mismatch (which introduces 4.1% strain). Dislocations are expected to increase phonon scattering and reduce thermal transport[1,35], and a similar reduction has been observed in GaN at room temperature for comparable dislocation densities[36].

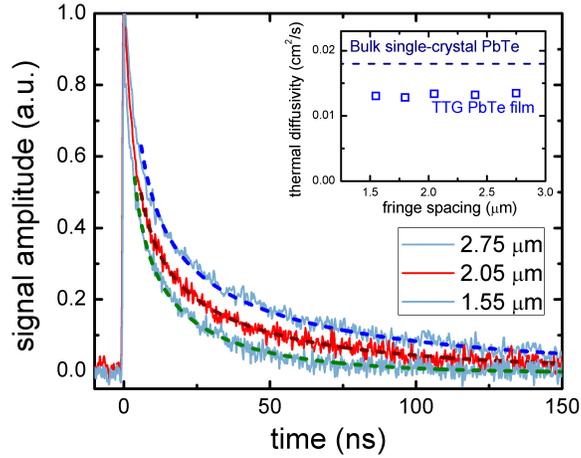

FIG. 5. (Color online) Amplitude grating signal for PbTe film for three representative transient grating periods (1.55μm, 2.05μm, and 2.75μm). The darker dashed curve with each trace is the best-fit using the 2-D diffusion equation (Eq. 16).

The thermal grating period was always smaller than the film thickness of 5 μm. Therefore, the presence of the BaF$_2$ substrate should have no effect on the analysis. Indeed, assuming a cross-plane thermal diffusivity equal or close to the measured in-plane diffusivity, the thermal penetration depth 2($\alpha t$)$^{1/2}$, where $t$=150 ns is the time window of the experiment, should only be ~1 μm, substantially shorter than the PbTe film thickness.

## VI. Modeling Phase Grating Signal

Even though we chose to use the amplitude grating signal for the thermal diffusivity measurements, the treatment of the subject matter would be incomplete without demonstrating that our model provides an adequate description of the phase grating signal as well. In particular, the initial dynamics of the phase signal poses an intriguing question. Both thermoreflectance and displacement contributions to the TG signal are expected to reach their maximum magnitudes shortly following the excitation pulse at *t*=0 and decay thereafter. Yet, the phase grating signal initially rises and reaches its maximum at *t*~4ns.  This behavior can be explained if we assume that thermoreflectance and displacement contribute to the phase grating signal with opposite signs.

According to Eq. 9, the phase grating signal is governed by the combination of thermoreflectance and displacement, $r''(t)-2k_p u(t)\cos\beta_p$.  Assuming a small absorption depth, $1/\zeta \to 0$, which is a good approximation for PbTe at 515 nm excitation wavelength, neglecting acoustic oscillations, and assuming isotropic thermal diffusivity, the surface displacement dynamics are given by[21]

$$u(t) = A_u \mathrm{erfc}\left(q\sqrt{\alpha t}\right) \ . \tag{18}$$

Fig. 6 shows the phase grating signal from Fig. 3 together with calculated traces using the thermal diffusivity determined from the amplitude grating data.  The dash-dotted line shows the pure surface-displacement response according to Eq. 18 and the dotted line shows the pure thermal response according to Eq. 17.  The solid dark line, which has excellent agreement with the phase grating data, is a linear combination of the calculated displacement and thermal responses, taken with opposite signs.  Thus the phase grating signal follows neither displacement nor temperature dynamics alone.  To employ this signal for the quantitative analysis of the thermal transport, one would need to use an additional fitting parameter, i.e., the ratio of the transient reflectivity and displacement contributions to the phase grating signal.  Such analysis might prove useful for samples yielding a weak amplitude grating signal and a strong phase signal.

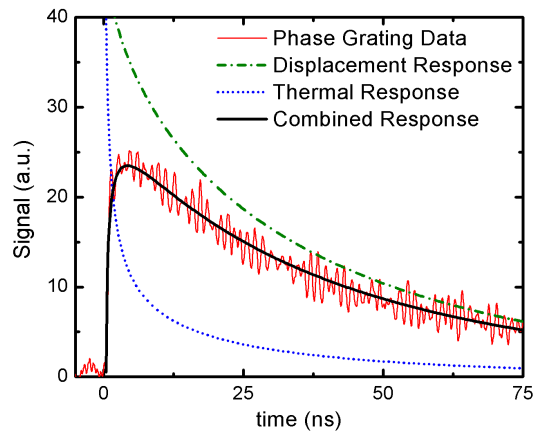

FIG. 6. (Color online) Phase-grating signal for 2.05 μm transient grating period.  The dash-dotted and dotted lines are respectively the calculated surface-displacement and thermal responses.  The solid dark line, the surface-displacement response minus the thermal response, represents the combined dynamics of the phase-grating signal.

## VII. Discussion and Conclusion

The presented methodology removes ambiguities in the analysis of transient thermal grating responses and lays the foundation for the use of the heterodyned transient grating method for thermal transport measurements in opaque materials. We have shown that ascribing the TG signal to the surface displacement alone may lead to significant errors in the analysis. By employing phase-controlled heterodyne detection, we have isolated the thermoreflectance component of the TG response and performed analysis with a two-dimensional thermal diffusion model to extract the in-plane thermal diffusivity of the sample.

We have taken advantage of a number of aspects of this technique and we can envision other benefits that could be utilized in future experiments. Unlike many transient thermoreflectance measurements[10,33,37,38], the method does not require a metal transducer film deposited on the sample surface. Another benefit of TG measurements is the ability to impart a well-defined length scale for thermal transport (effectively one half of the transient grating period $L$), which can be varied typically in the range 0.5-100 μm. Measurements at small periods are useful for characterization of thin films and on small areas (just a few TG fringes need to fit in an area to enable a measurement), while the ability to vary $L$ could be exploited to determine depth-dependent thermal properties[21,39]. On the other hand, the ability to control the thermal transport length scale opens prospects for studying non-equilibrium thermal transport at small distances. Measurable deviations from the Fourier law at room temperature are expected to occur, for example in gold at $L\sim1$ μm [40], and the same may be true for materials with high lattice thermal conductivity such as Si[41]. Reducing the grating period by using shorter optical wavelengths, possibly combined with other approaches such as immersion lens or near-field imaging, will open the avenue for studying thermal transport at the nanoscale[1]. The use of extreme ultraviolet (EUV) light to probe photothermal and photoacoustic responses has already been demonstrated[42]; using EUV to excite transient gratings would represent a great advance for nanometer range measurements albeit implementing phase-controlled heterodyne detection with EUV will be a challenge. We anticipate interesting further developments in both experimental methodology and applications to material science and thermal transport physics.